%% file: radcor00.tex
\documentclass[12pt]{article}
\usepackage{epsfig}

\newcommand\pubnumber{UCD-01-05}
\newcommand\pubdate{\today}
\newcommand\hepnumber{hep-ph/0106154}

\def\Re{{\rm Re}}
\def\Im{{\rm Im}}
\def\vevd{v_\Delta}
\def\what{\widehat}
\def\hhat{\what h}
\def\mhhat{m_{\hhat}}
\def\beq{\begin{equation}}
\def\eeq{\end{equation}}
\def\bit{\begin{itemize}}
\def\eit{\end{itemize}}

\def\del{\delta}

\def\mx{M_X}

\def\mz{m_Z}

\def\hi{h_i^0}
\def\mhi{m_{\hi}}

\def\h{h}
\def\mh{m_{\h}}

\def\mhmax{\mh^{\rm max}}
\def\mhmin{\mh^{\rm min}}
 
\def\lam{\lambda}

\def\wpm{W^{\pm}}
\def\hpm{H^{\pm}}

\def\call{{\cal L}}

\def\wtil{\widetilde}
\def\what{\widehat}
\def\tauptaum{\tau^+\tau^-}

\def\bfbm{\bf\boldmath}

\def\lsim{\mathrel{\raise.3ex\hbox{$<$\kern-.75em\lower1ex\hbox{$\sim$}}}}
\def\gsim{\mathrel{\raise.3ex\hbox{$>$\kern-.75em\lower1ex\hbox{$\sim$}}}}
\def\ifmath#1{\relax\ifmmode #1\else $#1$\fi}

\def\vev#1{\langle #1 \rangle}
\def\lam{\lambda}

\def\mplanck{M_{\rm P}}

\def\mhi{m_{h_1^0}}

\def\gl{\wt g}

\def\etc{{\em etc.}}

\def\dmm{\Delta^{--}}

\def\mdmm{m_{\dmm}}
\def\hdmm{h^{\dmm}}
\def\dpp{\Delta^{++}}
\def\delp{\Delta^{+}}

\def\hzero{\Delta^0}

\def\stop{\wt t}

\def\mstop{m_{\stop}}

\def\sbot{\wt b}

\def\msusy{m_{\rm SUSY}}

\def\gl{\wt g}

\def\hsm{h_{\rm SM}}
\def\mhsm{m_{\hsm}}
\def\hl{h^0}
\def\hh{H^0}
\def\ha{A^0}
\def\hp{H^+}
\def\hm{H^-}
\def\hpm{H^{\pm}}
\def\mhl{m_{\hl}}
\def\mhh{m_{\hh}}
\def\mha{m_{\ha}}

\def\mhpm{m_{\hpm}}
\def\tanb{\tan\beta}
\def\cotb{\cot\beta}

\def\mz{m_Z}
\def\mw{m_W}
\def\mgut{M_U}
\def\mx{M_X}

\def\wpm{W^{\pm}}

\def\cnone{\wt\chi^0_1}

\def\mcnone{m_{\cnone}}

\def\wt{\widetilde}

\def\emem{e^-e^-}
\def\mummum{\mu^-\mu^-}

\def\lam{\lambda}
\def\br{B}
\def\tauptaum{\tau^+\tau^-}

\def\gam{\gamma}

\def\anti{\overline}
\def\epem{e^+e^-}
\def\mupmum{\mu^+\mu^-}
\def\zstar{Z^\star}
\def\wstar{W^\star}

\def\mupmum{\mu^+\mu^-}

\def\rts{\sqrt s}
\def\ie{{\it i.e.}}

\def\eps{\epsilon}
\def\anti{\overline}

\def\gamhsm{\Gamma_{\hsm}^{\rm tot}}

\def\fbi{~{\rm fb}^{-1}}

\def\abi{~{\rm ab}^{-1}}

\def\gev{~{\rm GeV}}
\def\tev{~{\rm TeV}}

\def\hi{\h_1}
\def\hii{\h_2}

\def\mhi{m_{\hi}}

\def\dmm{\Delta^{--}}
\def\mdmm{m_{\dmm}}
\def\hdmm{h^{\dmm}}
\def\dpp{\Delta^{++}}

\def\hzero{\Delta^0}

\def\mhi{m_{h_1^0}}

\def\gl{\wt g}

\def\dmm{\Delta^{--}}

\def\mdmm{m_{\dmm}}
\def\hdmm{h^{\dmm}}
\def\dpp{\Delta^{++}}
\def\delp{\Delta^{+}}

\def\hzero{\Delta^0}

\def\emem{e^-e^-}
\def\dmm{\Delta^{--}}
\def\mdmm{m_{\dmm}}

\def\dpp{\Delta^{++}}
\def\delp{\Delta^{+}}

\def\hzero{\Delta^0}

\def\dmm{\Delta^{--}}
\def\mdmm{m_{\dmm}}
\def\hdmm{h^{\dmm}}
\def\dpp{\Delta^{++}}

\def\hzero{\Delta^0}

\def\lam{\lambda}
\def\anti{\overline}

\def\Title#1{\begin{center} {\Large\bf #1 } \end{center}}
\def\Author#1{\begin{center}{ \sc #1} \end{center}}
\def\Address#1{\begin{center}{ \it #1} \end{center}}

\newcommand\pubblock{\rightline{\begin{tabular}{l} \pubnumber\\
         \pubdate\\ \hepnumber \end{tabular}}}
\newenvironment{Abstract}{\begin{quotation}  }{\end{quotation}}
\newenvironment{Presented}{\begin{quotation} \begin{center} 
             Presented at the\end{center}
      \begin{center}\begin{large}}{\end{large}\end{center} \end{quotation}}

\makeatletter
\def\section{\@startsection{section}{0}{\z@}{5.5ex plus .5ex minus
 1.5ex}{2.3ex plus .2ex}{\large\bf}}
\def\subsection{\@startsection{subsection}{1}{\z@}{3.5ex plus .5ex minus
 1.5ex}{1.3ex plus .2ex}{\normalsize\bf}}
\def\subsubsection{\@startsection{subsubsection}{2}{\z@}{-3.5ex plus
-1ex minus  -.2ex}{2.3ex plus .2ex}{\normalsize\sl}}

\renewcommand{\@makecaption}[2]{%
   \vskip 10pt
   \setbox\@tempboxa\hbox{\small #1: #2}
   \ifdim \wd\@tempboxa >\hsize     
       \small #1: #2\par          
     \else                        
       \hbox to\hsize{\hfil\box\@tempboxa\hfil}
   \fi}

 \def\citenum#1{{\def\@cite##1##2{##1}\cite{#1}}}
 
\newcount\@tempcntc
\def\@citex[#1]#2{\if@filesw\immediate\write\@auxout{\string\citation{#2}}\fi
  \@tempcnta\z@\@tempcntb\m@ne\def\@citea{}\@cite{\@for\@citeb:=#2\do
    {\@ifundefined
       {b@\@citeb}{\@citeo\@tempcntb\m@ne\@citea\def\@citea{,}{\bf ?}\@warning
       {Citation `\@citeb' on page \thepage \space undefined}}%
    {\setbox\z@\hbox{\global\@tempcntc0\csname b@\@citeb\endcsname\relax}%
     \ifnum\@tempcntc=\z@ \@citeo\@tempcntb\m@ne
       \@citea\def\@citea{,}\hbox{\csname b@\@citeb\endcsname}%
     \else
      \advance\@tempcntb\@ne
      \ifnum\@tempcntb=\@tempcntc
      \else\advance\@tempcntb\m@ne\@citeo
      \@tempcnta\@tempcntc\@tempcntb\@tempcntc\fi\fi}}\@citeo}{#1}}
\def\@citeo{\ifnum\@tempcnta>\@tempcntb\else\@citea\def\@citea{,}%
  \ifnum\@tempcnta=\@tempcntb\the\@tempcnta\else
  {\advance\@tempcnta\@ne\ifnum\@tempcnta=\@tempcntb \else\def\@citea{--}\fi
    \advance\@tempcnta\m@ne\the\@tempcnta\@citea\the\@tempcntb}\fi\fi}
\makeatother

\input econfmacros2.tex
\def\csumb{Davis Institute for High Energy Physics, Department of Physics\\
University of California at Davis, Davis CA, USA}
\def\support{\footnote{Work supported in part by the
U.S. Department of Energy and by the Davis Institute for
High Energy Physics.}}

\begin{document}
\begin{titlepage}
\pubblock

\vfill
\def\thefootnote{\fnsymbol{footnote}}
\Title{Hunting the Higgs Boson(s)}
\vfill
\Author{John F. Gunion \support}
\Address{\csumb}
\vfill
\begin{Abstract}
I give a brief review of some of the opportunities and challenges
that could arise in our quest to unravel the Higgs sector
that very probably underlies electroweak symmetry breaking.
In particular, I review scenarios with an extended Higgs sector that 
allow for a heavy SM-like Higgs boson and/or make
discovery more difficult while at the same time maintaining
consistency with current limits and precision electroweak constraints.
\end{Abstract}
\vfill
\begin{Presented}
5th International Symposium on Radiative Corrections \\ 
(RADCOR--2000) \\[4pt]
Carmel CA, USA, 11--15 September, 2000 \\
{\normalsize and at} \\
8th International Symposium on Particles, Strings and Cosmology\\
(PASCOS--2001) \\[4pt]
University of North Carolina, Chapel Hill, 10--15 April, 2001\\ 
\end{Presented}
\vfill
\end{titlepage}
\def\thefootnote{\arabic{footnote}}
\setcounter{footnote}{0}

\section{The Standard Model}

The most important and immediate goal in our quest
to understand nature at the microscopic level is the determination
of the mechanism by which elementary
particles acquire mass. One very attractive approach is to
hypothesize the existence of a Higgs sector (for a review, see \cite{hhg})
of scalar fields (some of which must have non-zero
quantum numbers under the weak SU(2)$\times$U(1)
electroweak gauge group). The Higgs potential must be such that one
or more of the neutral components of the Higgs fields spontaneously acquires 
a non-zero
vacuum expectation value, thereby giving masses to the $\wpm$ and $Z$
gauge bosons. In the minimal Standard
Model (SM), mass generation is accomplished through the existence
of a Higgs sector containing a single complex
scalar field doublet (under weak isospin), $\phi=\left(\begin{array}{c}
\phi^+ \cr \phi^0\end{array}\right)$.
When $\Re \phi^0$ 
acquires a vacuum expectation value ($v\over\sqrt2$), the $\phi^\pm$
and $\Im \phi^0$ fields are absorbed
by the hitherto massless $W^\pm$ and $Z$ fields which thereby acquire mass.
At the same time, Yukawa couplings $\lam_f\anti f f \phi$ lead
to the generation of mass for the fermions, $m_f\propto \lam_f v$.
The quantum fluctuations of the remaining field $\Re \phi^0$, 
correspond to a physical particle, the neutral Higgs boson, denoted $\hsm$. 
The couplings of the $\hsm$ to other SM particles are completely constrained.
However, the mass of the $\hsm$ is completely unconstrained in
the SM context without referencing physics at higher energy scales.

\begin{figure}[h]
\leavevmode
\vskip -.3in
\epsfxsize=4in
\centering
\epsffile{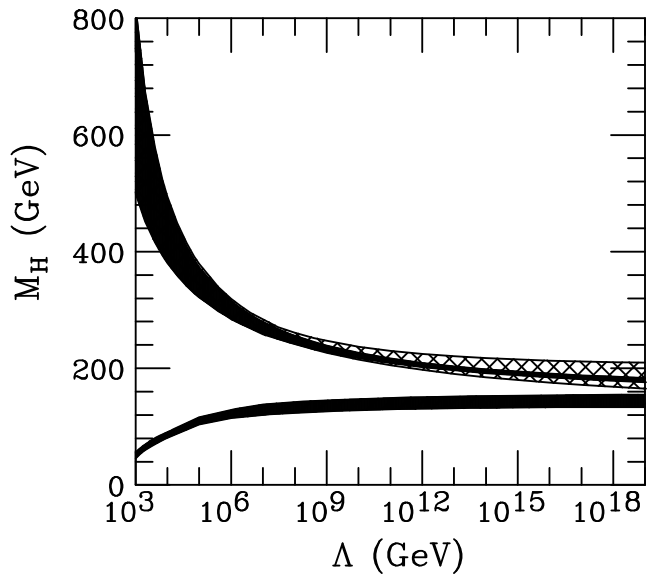}
\caption{Triviality and (meta)stability bounds for the SM
Higgs boson as a function of the new physics scale $\Lambda$.
From \protect\cite{hambye}.}
\label{smbounds}
\end{figure}
If the SM is the correct description of electroweak symmetry breaking
at current energies, it could still be
that the SM is only an effective theory valid below some
higher energy scale $\Lambda$.  Above $\Lambda$, new physics enters
and a more complete/fundamental theory would emerge. One
possibility is that there is no new physics between electroweak scales
and the Planck scale, $\mplanck$. Or it could be that a theory
such as supersymmetry emerges at a lower scale.  
Fig.~\ref{smbounds} (from \cite{hambye})
shows that the SM could remain valid as an effective
theory all the way up to $\mplanck$ only for a very limited range of $\mhsm$,
roughly $140<\mhsm<180\gev$.  For $\mhsm$ outside this range,
new physics would have to enter at a much lower scale.  For example,
if a $\mhsm\sim 115\gev$ SM Higgs boson is discovered, 
then $\Lambda\lsim 1000\tev$.
The upper bound shown in Fig.~\ref{smbounds} derives from
requiring that the coupling
$\lam$ appearing in the Higgs field quartic self-coupling term in the 
Higgs potential, $\propto \lam\phi^4$, remain perturbative
when `probed' at energy scale $\Lambda$.
Since $\lam$ grows with energy scale,
this bounds $\lam(\mhsm)$, thereby bounding $\mhsm\sim 2v^2\lam(\mhsm)$.
The lower bound shown derives from requiring stability of
the potential. In particular, $\lam$ is not allowed to be driven negative
at energy scales below $\Lambda$ 
(by the large top quark contribution to the running of $\lam$);
\ie\ we require $\lam(\Lambda)>0$. Without this constraint the universe
would ultimately prefer to tunnel to a state in which the Higgs field $\phi$
has values with $|\phi|\gsim\Lambda$, yielding large negative 
$V(\Lambda)\propto \lam(\Lambda)|\phi\sim\Lambda|^4$ if $\lam(\Lambda)<0$.
The meta-stability condition, that the time
scale for such tunneling be longer than the age of the universe,
is only slightly less constraining.

\begin{figure}[b!]
\vskip -2in
\epsfxsize=5in
\centering
\hspace*{.3in}\epsffile{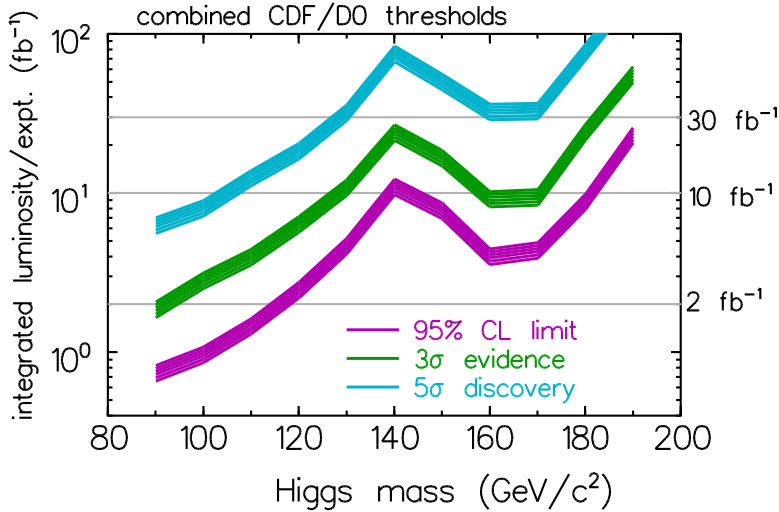}
\vskip -2in
\caption{Prospects for SM Higgs discovery at the Tevatron.}
\label{final}
\end{figure}

Precision electroweak data
suggests \cite{deltaral2} the presence of a light SM-like $\h$, the best 
single SM-like Higgs boson fit being
obtained for $\mh\lsim 100\gev$. 
Recent LEP data \cite{igo} contain hints (at the roughly 2.9$\sigma$ level)
that a SM-like Higgs boson might be present with mass $\mh\sim 115\gev$. 
This same data could also be interpreted as providing weak evidence
for a somewhat spread-out Higgs signal in the region $\mh\lsim 115\gev$, 
such as might arise if there were a number of Higgs
bosons with overlapping resonance shapes, each one having $ZZ$
coupling-squared that is a small fraction of the strength expected
for the $\hsm$.

If the precision electroweak and LEP hints for a single
light SM-like Higgs boson are correct,
the Tevatron will have an excellent chance of detecting such an $\h$
with $L=15\fbi$ of accumulated luminosity (per experiment).
This is illustrated in Fig.~\ref{final}, from \cite{tevreport}.
If $90~{\rm GeV}~\lsim\mhsm\lsim 130$~GeV one employs 
$q'\anti q\to W^*\to W\hsm\to \ell\nu b\bar b$ 
and $q\anti q\to Z^*\to Z\hsm\to\nu\bar\nu b\bar b$ 
and $\ell^+\ell^-b\bar b$.
If $130\gev\lsim \mhsm\lsim 190\gev$ one uses 
$gg,W^*W^*\to \hsm$ as well as
$q'\anti q\to W^*\to W\hsm$ and $q\anti q\to Z^*\to Z\hsm$, all with
$\hsm\to WW^*,ZZ^*$. Relevant final states for $\hsm$ decay
would be $\ell^\pm\ell^\pm jjX$ and $\ell^+\ell^-\nu\bar\nu$.
Currently, it is believed that $L=15\fbi$ can be accumulated
by 2006-2007, i.e. just as LHC will start producing physics results.

\begin{figure}[h!]
\centering 
\hspace*{.6in}\psfig{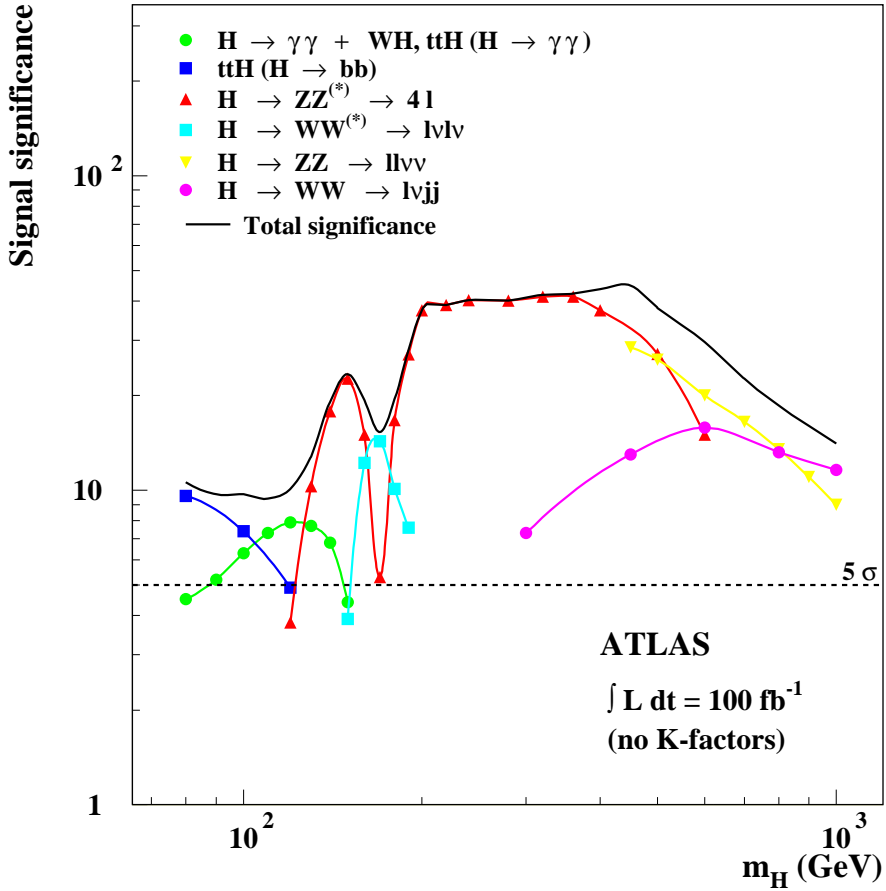}
\vspace{.15in}
\caption[0]{The statistical significance in various channels
for a Standard Model Higgs
signal with $L=100\fbi$ of accumulated luminosity for the ATLAS detector
at the LHC. Also shown is the net statistical significance after
combining channels. From \cite{atlashsmref}. The CMS
detector finds similar results \cite{cmshsmref}.}
\label{atlashsm}
\end{figure}

The LHC and its detectors have been specifically designed
to discover the $\hsm$ for any $\mhsm\lsim 1\tev$ or to see signs
of a strongly interacting $W$ sector if the effective Higgs mass
is even larger. The discovery modes are the following.
For $\mhsm\lsim 130\gev$, one employs 
$gg,W^*W^*\to\hsm\to\gam\gam$, $q_i\anti q_j\to W^\pm \hsm$
and $gg\to t\anti t\hsm$ with $\hsm\to\gam\gam$ and $\hsm\to b\anti b$.
For $\mhsm>130\gev$, the best
signal is $gg,W^*W^*\to \hsm \to ZZ^{(*)}\to 4\ell$ 
($gg,W^*W^*\to\hsm\to WW^*\to \ell\nu\ell\nu$ for $\mhsm\sim 2\mw$).
If $\mhsm>300\gev$ ($400\gev$) the 
$gg,W^*W^*\to\hsm\to WW\to \ell\nu jj$ ($\to ZZ\to \ell\ell\nu\nu$) modes
are very robust. The statistical significances
for various channels are shown in Fig.~\ref{atlashsm}. For $L=100\fbi$,
a signal of at least $10\sigma$ is achieved for all $\mhsm <1\tev$.

A future linear $\epem$ collider would also be
certain to detect the SM $\hsm$ unless $\mhsm>\rts$. 
Comprehensive reviews are found in \cite{teslareport}
and \cite{orangebook}. If $\mhsm<\rts-\mz$,
$\epem\to Z^*\to Z\hsm$ production would allow
both an inclusive recoil mass determination of $\sigma(Z\hsm)$
and exclusive final state determinations of $\sigma(Z\hsm)\br(\hsm\to X)$
for various final states $X$. The ratio of the latter to the former
gives a result for $\br(\hsm\to X)$. 
The power of this approach and of the LC detectors
to separate the various channels $X$ is illustrated
by the fact that for $L=500\fbi$ of
accumulated luminosity (1 or 2 years of operation) one can even obtain
an accurate determination of $\br(\hsm \to WW^*)$ if $\mhsm\gsim 120\gev$
and of $\br(\hsm\to \gam\gam)$ if $\mhsm\lsim 130\gev$.
Even the very narrow width of the light Higgs
can be determined quite accurately by indirect means.  For instance,
by isolating  $\epem\to\epem W^*W^*\to\epem \hsm$ events one can extract
$\gamhsm={\sigma(W^*W^*\to\hsm\to WW^*)\over [\br(\hsm\to WW^*)]^2}$.
The $\gam\gam$ collider option at the LC can also play
an important role.  In particular, the $\gam\gam\to \hsm$
coupling can be determined from the ratio
$\sigma(\gam\gam\to\hsm\to b\anti b)/\br(\hsm\to b\anti b)$ 
(the latter determined using the $Z\hsm$ techniques).
This coupling is very sensitive to the presence
of loops containing heavy particles whose mass is acquired
via the Higgs mechanism.  In addition, at low masses such that
the $W^*W^*$ technique for total width determination is not
very accurate, 
$\gam\gam\to\hsm$ allows \cite{Gunion:1996qg} extraction of 
$\gamhsm={\sigma(\gam\gam\to\hsm\to b\anti b)\over \br(\hsm\to\gam\gam)
\br(\hsm\to b\anti b)}$.
The process $\gam\gam\to\hsm$ also allows determination of the CP 
nature of the $\hsm$ by studying the cross section
dependence upon relative orientation of the (transverse) 
polarizations of the colliding 
$\gam$'s \cite{Grzadkowski:1992sa,Gunion:1994wy,Kramer:1994jn}. 
CP=+ ($-$) implies $\gam\gam\to\hsm$ cross section proportional to 
$\vec \eps_1\cdot \vec \eps_2$
($\vec\eps_1\times \vec \eps_2$).
Finally, by studying angular distributions of the $t$, $\anti t$ and $\h$
in $\epem\to t\anti t\h$ it is possible to determine the CP of the resonance 
eigenstate \cite{Gunion:1996bk,Gunion:1996vv}.

\section{Non-Exotic Extensions of the SM Higgs Sector}

Even within the SM effective field theory context, the Higgs sector
need not consist of just a single doublet;
one should consider extended Higgs sector
possibilities. Indeed, typical models in which the Higgs
sector is the result of a new strong interaction at a higher
scale $\Lambda$ produce an effective field theory
below $\Lambda$ that contains at least two doublets and/or extra 
singlets \cite{Dobrescu:1999cs}. 
Higher representations are also a possibility. 
String models also often yield quite a number of 
Higgs representations at low energy \cite{Cvetic:2000nc}; 
singlets, doublets and
higher representations are all possible.

Addition of singlets
poses no particular theoretical problems (or benefits).
Addition of one or more extra doublet representation(s)
has both attractive and unattractive aspects. On the unattractive
side is the fact that 
the squared-mass(es) of the additional charged Higgs boson(s) 
become new parameter(s)
that must be chosen to be positive definite in order to avoid breaking
of electromagnetic symmetry. This unfavorable aspect is, in the view
of many, more than compensated by the fact that a multi-Higgs-doublet
model allows for the possibility of explaining all CP-violation
phenomena as a result of 
explicit or spontaneous CP violation in the Higgs sector.
Triplet representations and higher are deemed `exotic' in that
$\rho$ is no longer computable when they participate
in EWSB (\ie\ when the vev of the neutral member of the representation
is non-zero); instead,
$\rho$ becomes infinitely renormalized and must be treated
as an input parameter to the model \cite{Gunion:1991dt}.
In this section, we focus our attention on singlet and doublet extensions.
In both cases, detection and simulation considerations change dramatically.
Triplets will be discussed very briefly in the next section.

The new considerations that arise for an
extended SM Higgs sector are brought most immediately into focus
by discussing the discovery prospects for Higgs bosons at
an $\epem$ collider; other colliders will encounter even
greater difficulty in ensuring discovery of at least one Higgs boson
of an extended sector.

\subsection{A Continuum Signal}
As stated above, it is not entirely unreasonable to consider a case in which
there are many singlets and/or
extra doublets, possibly even triplets,
in addition to the original doublet Higgs field $\phi$.
Each complex neutral field results in an extra scalar and extra pseudoscalar
degree of freedom. The former will generally mix with $\Re\phi^0$ and
the interesting question is the extent to which this could make Higgs
discovery difficult.  The worst case scenario is that in which
the physical eigenstates 
share the $WW/ZZ$ coupling-squared and are spread out in mass
in such a way that their separation is smaller than the $\sim 10\gev$
or so mass resolution of the detector.
The result could be a very spread out and diffuse signal that
could only be searched for as a broad excess in the $\mx$ recoil
mass spectrum in $\epem\to ZX$ production, where 
$\mx$ is computed from $p_X=p_{e^+}+p_{e^-}-p_Z$
for events in which the $Z$ decays to $\epem$, $\mupmum$
(and possibly jets).
As noted earlier, LEP2 data is consistent with 
a small spread-out excess of events
at high $\mx$ (in the $\mx\sim 100-110\gev$ region)
beyond the number predicted by background computations; this excess
could be interpreted in terms of such a diffuse
spread-out signal.

Fortunately, there are constraints on this scenario. First, defining $C_i$
to be the strength of the $h_iVV$ coupling relative to 
that of the $\hsm$, unitarity for $WW$ scattering, as well as the general
structure of the theory, imply the
sumrule $\sum_i C_i^2\geq1$; if only singlet
and doublet representations are present $\sum_iC_i^2=1$.  
Second, precision electroweak constraints
imply that the value of $\vev{M^2}$ appearing in
\beq
\sum_i C_i^2m_{h_i}^2=\vev{M^2}\,.
\label{msqlim}
\eeq
should not exceed about $(200\gev)^2$. For the most general
supersymmetric model Higgs sector, imposing
the requirement of perturbativity of couplings
after evolving up to the GUT
scale  yields this same result for the maximum possible 
$\vev{M^2}$ \cite{Espinosa:1998re}.

To illustrate the consequences \cite{Espinosa:1999xj}, 
assume $C_i^2$ is constant from $\mhmin$ to $\mhmax$;
using continuum notation, $C^2(\mh)\geq 1/(\mhmax-\mhmin)$
for $\int d\mh C^2(\mh)\geq 1$, while Eq.~(\ref{msqlim})
implies ${1\over 3}([\mhmax]^2+\mhmax\mhmin+[\mhmin]^2)\leq \vev{M^2}$. 
Let us also suppose
that LEP data can be used to show that $C^2(\mh)$ is very small
for $\mh<70\gev$ (this is being examined currently). Then
if $C^2(\mh)$ is constant from $\mhmin=70\gev$
out to $\mhmax=300\gev$ the sumrule will be saturated.

Clearly, LEP2 would have had great difficulty confirming the presence
of such a broad excess. One needs to have $\epem$ collisions
at high enough energy to avoid kinematic suppression over the bulk
of the $\mx$ region in question.  A $\rts=500\gev$ collider would
be more or less ideal. In Ref.~\cite{ghs}, the backgrounds
in the recoil $\mx$ spectrum for $ZX$ production were examined
for $\rts=500\gev$ over the interval $70\gev$ to $200\gev$.
For the $\mhmin=70,\mhmax=300\gev$ case described,
a fraction $f\sim 0.57$ of the continuum signal resides  in this region.
In order to avoid the large $ZZ$ background, it is actually best
to restrict consideration to the $100-200\gev$ range in which a
fraction $f\sim 0.43$ of the signal resides. For $L=500\fbi$,
the excess signal event rate after cuts
would be $S\sim 1350 f\sim 580$ with a background
of $B=2700$. The resulting 
$\sim 50\%f\sim 22\%$ excess over background would be readily detected,
and would yield  $S/\sqrt B\sim 26f\sim 11$.
Allowing for some extra weighting of the signal into the $\mx=\mz$
region, it still seems safe to say that $S/\sqrt B>5$ would be
achieved for $L\gsim 200\fbi$.

Obviously, detection of this type of signal would be very difficult,
if not impossible, at a hadron collider due to the inability
to reconstruct the recoil mass in a $ZX$ or $WX$ event (the
energies of the colliding quarks being unknown).
In this scenario, the LHC would have good evidence
that $WW$ scattering at high-$m_{WW}$ 
was perturbative, but the continuum of Higgs bosons
responsible for this perturbativity would probably not be directly detected.
Detection of the Higgs bosons would require that
only a few of the Higgs bosons decayed to some particular identifiable
final state (e.g. $b\anti b$) and that these same Higgs bosons 
were sufficiently well separated
in mass that the individual mass peaks could be reconstructed.
This latter is a possibility if some of the Higgs bosons give
mass to some fermions and not others, rather than all Higgs bosons
contributing roughly equally to the various fermionic masses.
Of course, this type of channel separation would make resonance
peak reconstruction possible at the LC as well.

\subsection{The General Two-Higgs-Doublet Model}

This is a particularly useful model to consider since it already
displays many features that would be present in still more complex
Higgs sectors (see \cite{hhg} for a review and references).  
We will confine our attention to a type-II two-doublet
model (in which one Higgs doublet, $\phi_u$, gives mass to up quarks
while the second, $\phi_d$, gives mass to down quarks and leptons).
Of course, the MSSM Higgs sector is a constrained type-II two-doublet
model. If CP is conserved in the Higgs sector, then there
are two CP-even eigenstates, $\hl$ and $\hh$, one CP-odd
eigenstate, $\ha$ and a charged Higgs pair, $\hpm$.
If CP is violated, the $\hl,\hh,\ha$ would mix to form a trio
of mixed-CP eigenstates, $h_{i=1,2,3}$. One of the most
important parameters of a 2HDM model is $\tanb=v_u/v_d$,
the ratio of vacuum expectation values for the neutral
field components of $\phi_u$ and $\phi_d$; $v_u^2+v_d^2=v^2_{\rm SM}$
is required to obtain the correct $W$ and $Z$ masses.

The most pressing question is again whether or not we
are guaranteed that there should be at least
one light Higgs boson with properties such that we can detect it
at the various colliders.  The answer is model dependent.

As we shall review later, in the general 2HDM context (i.e.
without the constraints of supersymmetry), it is possible
to satisfy precision electroweak constraints even if the only
Higgs boson that has substantial $WW/ZZ$ coupling is quite
heavy (but, at most $\sim 1\tev$). Precision constraints are most easily
satisfied if there is one light Higgs boson (with no $WW/ZZ$ coupling),
all others being quite heavy. Would we discover this light $\hhat$?

Again we use the $\epem$ collider to illustrate.  There the relevant
production processes would be
$\epem\to t\anti t \hhat$, $\epem\to b\anti b \hhat$,
$\epem\to Z^*\to  Z \hhat\hhat$ and $\epem
\to \nu\nu \hhat\hhat$.  As regards the fermion processes,
there are sumrules that guarantee that the $b\anti b\hhat$
and $t\anti t\hhat$ couplings cannot both be 
suppressed \cite{Grzadkowski:2000wj}.
In particular, for a $\hhat$ of a general type-II 2HDM
with no $VV$ coupling one finds ${g_{t\anti t\hhat}/ g_{t\anti t\hsm}}=\cotb$
and ${g_{b\anti b\hhat}/ g_{b\anti b\hsm}}=\tanb$.
The $Z^*\to Z\hhat\hhat$ and $WW\to\hhat\hhat$ processes are dominated
by the quartic coupling which is determined purely by the covariant
gauge derivative structure, $(D_\mu \Phi)^\dagger (D^\mu \Phi)$,
responsible for the relevant interactions.
We will now outline why these processes are not
necessarily sufficient to guarantee $\hhat$ discovery.

\medskip
\noindent{\bfbm Yukawa processes}
\medskip

Because of the $\tanb$ dependence of the couplings, 
$\epem\to t\anti t\hhat$ will always yield an observable
event rate if $\tanb$ is small enough (and
the process is kinematically allowed) while $b\anti b \hhat$
will be observable for large enough $\tanb$.
\begin{figure}[h!]
\centering
\epsfxsize=6in
\epsffile{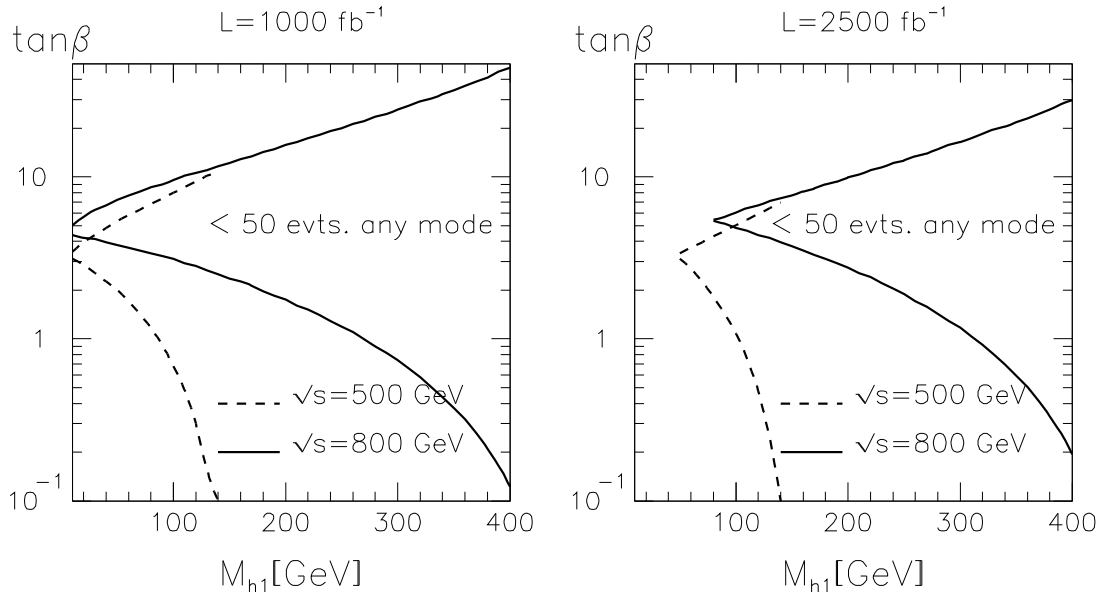}
\caption{
For $\protect\rts=500\gev$ (dashes) and $\protect\rts=800\gev$ (solid),
we plot the maximum and minimum $\tanb$
values between which $t\anti t \hhat$ and $b\anti b \hhat$ 
both have fewer than 50 events assuming
(a) $L=1000\fbi$ or (b) $L=2500\fbi$.}
\label{worstcaseband}
\end{figure}
However, even for $L=2500\fbi$ there is a wedge of $\tanb$, beginning
at $\mhhat\sim 50\gev$ ($80\gev$) for $\rts=500\gev$ ($800\gev$), 
for which neither process will have as many as 
50 events \cite{Grzadkowski:2000wj,Chankowski:2000an}, deemed
the absolute minimum number of events for which
detection would be possible at the LC. Of course, the upper limit
limit for the wedge
illustrated up to $400\gev$ rises further for still  larger $\mhhat$
values, while the lower line deliminating the wedge disappears
once $t\anti t \hhat$ is kinematically forbidden.
In short, the fermionic coupling sum rules do not
yield any guarantees. They only restrict the problematical region.

\medskip
\noindent{\bfbm Double Higgs production processes}
\medskip

\begin{figure}[t!]
\centering
\psfig{file=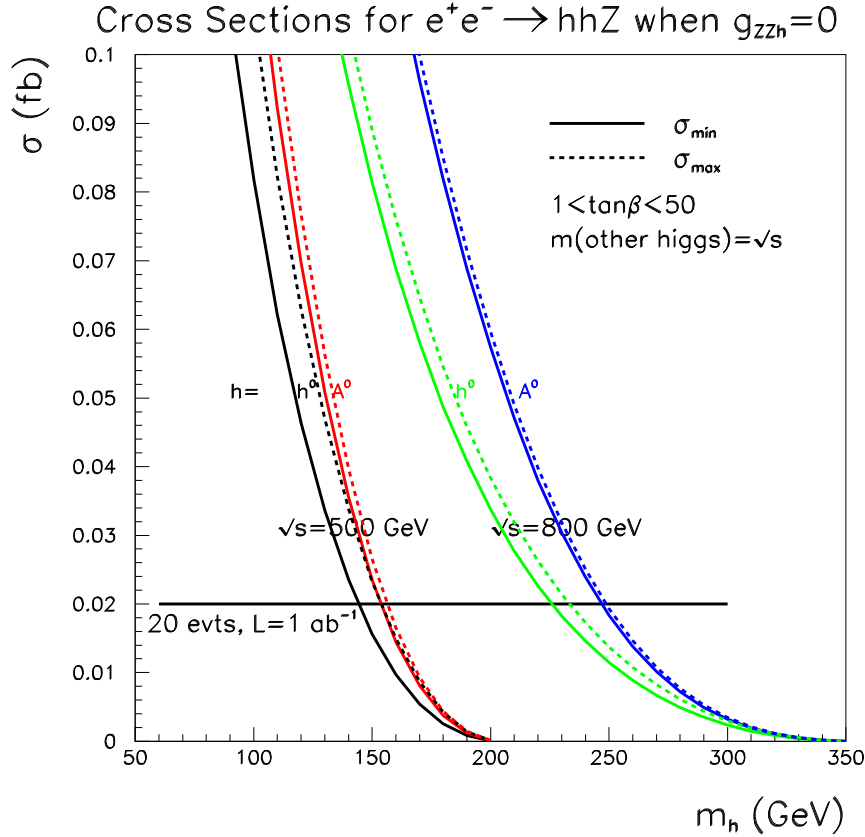,width=12cm}
\caption{
For $\protect\sqrt s=500\gev$  and $800\gev$ and for $\hhat=\hl$
and $\hhat=\ha$,
we plot as a function of $\mhhat$ the maximum and minimum values of
$\sigma(\epem\to \hat h\hat h Z)$  found after scanning $1<\tanb<50$
taking all other Higgs masses equal to $\sqrt s$. For $\hat h=\hl$, we
require $\sin(\beta-\alpha)=0$ during the scan. The 20 event level
for $L=1\abi$ is indicated.}
\label{f:hhz}
\end{figure}
In Fig.~\ref{f:hhz} we \cite{gunfarris} plot the cross section for $\epem\to
\zstar\to Z\hhat\hhat$. We see that this process can probe
up to $\mhhat=150\gev$ ($250\gev$) for a 20 event signal
with $L=1000\fbi$ (50 events for $L=2500\fbi$).
Similar results are obtained for $WW\to \hhat\hhat$ fusion production.
Thus, even after combining these process with the Yukawa processes,
there is a large range of $\mhhat$ and $\tanb$ values for which
the only Higgs boson light enough to be produced in $\epem$
collisions cannot be detected.

\medskip
\noindent{\bfbm Precision electroweak constraints in the 2HDM}
\medskip
\begin{figure}[t!]
\centering
\psfig{file=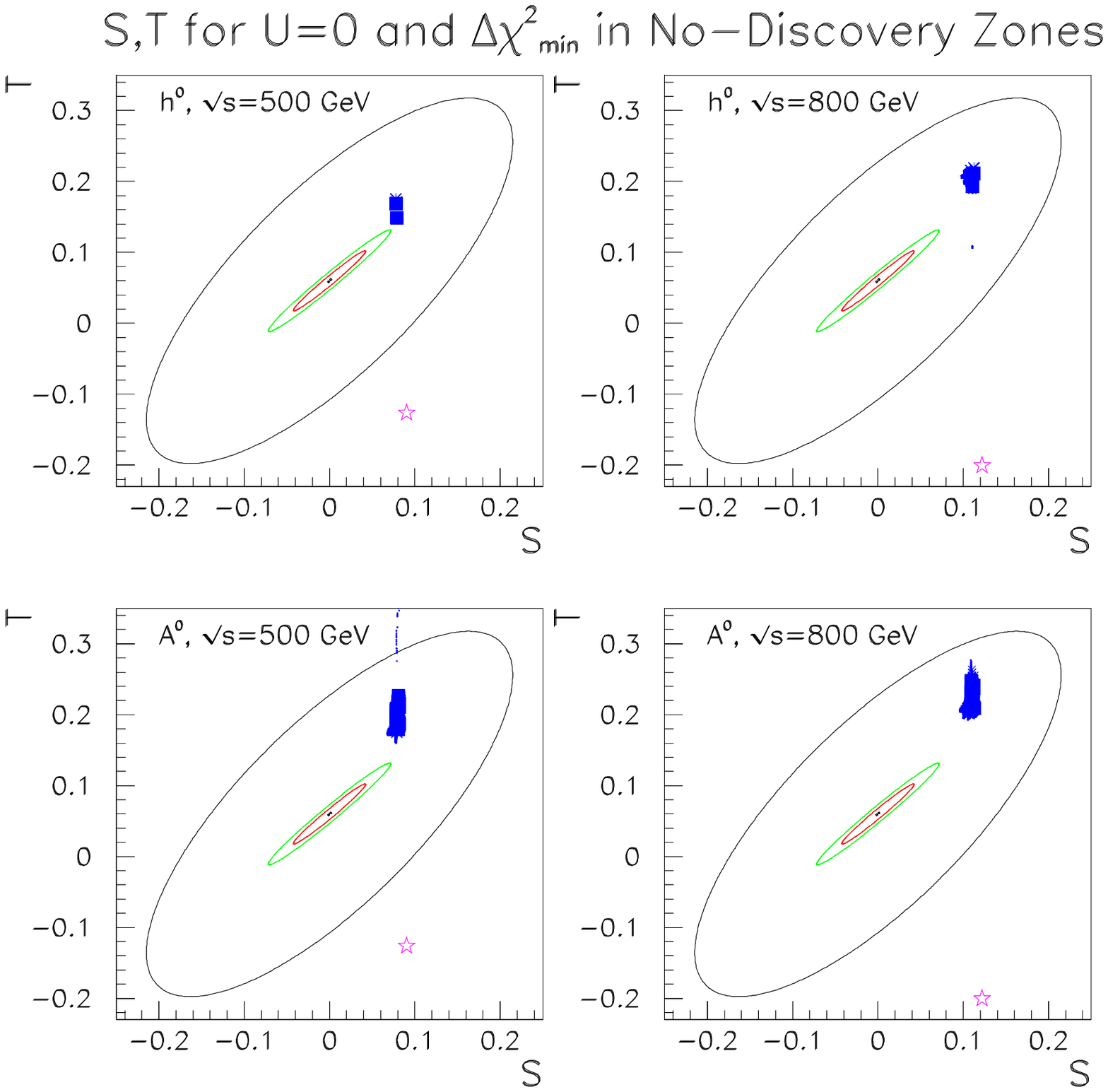,width=12cm}
\caption{The outer ellipses indicate the current precision
electroweak 90\% CL region in the $S,T$ plane
for $U=0$ and $\mhsm=115$ GeV. 
The innermost (middle) ellipse show the size of the 90\% (99.9\%) CL
region for $\mhsm=115$ GeV after new precision electroweak
measurements (especially of $\sin^2\theta_{\rm leptonic}$ at
a Giga-$Z$  factory {\it and} a $\Delta m_W\protect\lsim 6$ MeV threshold
scan measurement. 
The blobs indicate the $S,T$ predictions for points with $\tanb>2$
that lie within the no-discovery wedges illustrated in Fig.~\ref{f:hhz},
adjusting other model parameters so that the $\Delta\chi^2$
of the precision electroweak fit is minimized while keeping
all but one Higgs boson heavier than $\rts$.
Stars show $S,T$ predictions for the SM taking
$\mhsm=500$ or $800$ GeV.
} 
\label{dsdt}
\end{figure}

In this subsection, we demonstrate how a 2HDM can give good
agreement with precision electroweak constraints, even if
there is only one  Higgs boson with $VV$ decoupling and it has mass 
$\sim 1\tev$ \cite{Chankowski:2000an}. As noted earlier, a heavy SM-like $\h$ gives
a large $\Delta S>0$ and large $\Delta T<0$, as illustrated by the
locations of the stars in Fig.~\ref{dsdt}. The key is to compensate
the negative $\Delta T$ from the $1\tev$
SM-like Higgs with a large $\Delta T>0$ from a
small mass non-degeneracy (weak isospin breaking) of
heavier Higgs. For example, for a light $\hhat=\ha$, 
the $\hl$ would be taken heavy and SM-like and the value of
$\Delta\rho$ would be approximately given by: 
\beq
   \Delta \rho=\frac{\alpha}{16 \pi m_W^2 c_W^2}\left\{\frac{c_W^2}{s_W^2}
   \frac{\mhpm^2-\mhh^2}{2}-3m_W^2\left[\log\frac{\mhl^2}{m_W^2}
   +\frac{1}{6}+\frac{1}{s_W^2}\log\frac{m_W^2}{m_Z^2}\right]\right\}\nonumber
\label{drhonew}
\eeq
From this formula, it is clear that 
one can adjust $\mhpm-\mhh\sim {\rm few}\gev$ (both $\mhpm$ and $\mhh$
being large) 
so that the $S,T$ prediction moves to the location of the blobs
shown.

\medskip
\noindent{\bfbm Possible evidence from $a_\mu$ for a light $\hhat=\ha$}
\medskip

\begin{figure}[h!]
\centering\psfig{file=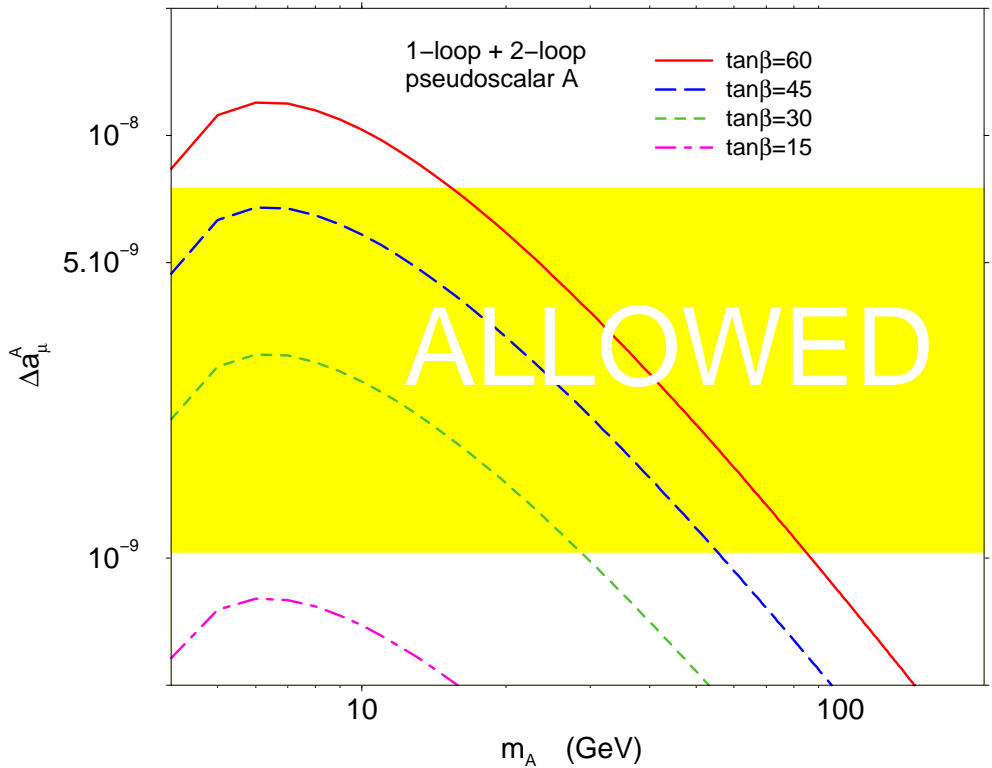,width=13cm}
\caption{Explanation of new BNL $a_\mu$ value via light 2HDM $\ha$. (Cheung,
Chou, Kong)}
\label{amucheung}
\end{figure}

The latest BNL result \cite{bnlresult} for $a_\mu$ differs by 2.6$\sigma$ from 
the SM prediction
(for a standard set of inputs for low energy $\sigma(\epem\to {\rm hadrons})$):
\begin{equation} \label{exp}
\Delta a_\mu \equiv a_\mu^{\rm exp} -a_\mu^{\rm SM} = 
426 (165) \; \times \; 10^{-11} \;.
\end{equation}
Taking the above numbers at face value, 
the range of $\Delta a_\mu$ at 95\% C.L. 
($\pm 1.96\sigma$) is given by
$10.3\times 10^{-10} < \Delta a_\mu < 74.9 \times 10^{-10} \;.$
A light $\ha$ ($\hl$) gives a positive (negative) contribution 
to $a_\mu$ dominated (for all but a very light Higgs boson)
by the two-loop Bar-Zee graph.  If we use a light $\ha$ as the
entire explanation for $\Delta a_\mu$, 
Fig.~\ref{amucheung} shows that this leads to constraints
such that $\tanb>15$ is required with $\mha<100\gev$ (smaller values
for smaller $\tanb$) \cite{amucheungref}.
For $\tanb>17$ and $\mha<100\gev$, the $\ha$
will be found at a LC for sure, but discovery of such a light
state primarily decaying into two (soft) jets will be
hard at the LHC. If the size of $\Delta a_\mu$ should decline
with analysis of the final data set, or with alternative
input for $\sigma(\epem\to {\rm hadrons})$ at low energy,
higher $\mha$ and/or smaller $\tanb$ would be needed to explain
$\Delta a_\mu$.  Thus, for smaller $\Delta a_\mu$ the 
$\ha$ might not be observable at either the LC or the LHC.
Of course, there are many other new-physics explanations for $\Delta a_\mu$.
Possibly a piece could come from the Higgs sector and a piece from
these other sources.
\vskip -.3in

\section{Triplet Representations}

It is certainly easy to construct models in which the Higgs 
sector contains one or more triplet representations
(see \cite{hhg} for a review of models).
Most interesting would be the presence of a complex $|Y|=2$
triplet representation.
One can use a $2\times 2$ matrix notation for such a representation:
\begin{equation}
\Delta=\pmatrix{\delp/\sqrt{2} & \dpp \cr \hzero & -\delp/\sqrt{2} \cr}\,.
\end{equation}
The most dramatic new features of Higgs representations containing
a complex triplet are the presence of the doubly-charged Higgs bosons, $\dmm$
and $\dpp$ and the possibility of lepton-number-violating Majorana-like
couplings which take the form  
\begin{equation}
{\cal L}_M=ih_{ij}\psi^T_{i} C\tau_2\Delta\psi_{j}+{\rm h.c.}
\,,
\label{couplingdef}
\end{equation}
where $i,j=e,\mu,\tau$ are generation indices.
This coupling will give rise to a see-saw mass term
if $\vev{\Delta^0}\equiv \vevd\neq 0$. 
However, if $\vevd\neq 0$, then we lose predictivity for $\rho$;
$\rho$ is renormalized and becomes an input parameter for the model
\cite{Gunion:1991dt}.
Whether or not $\vevd\neq 0$, $\call_M$ gives rise to 
$\emem\to\dmm$ and $\mu^-\mu^-\to \dmm$ couplings.
Left-right (L-R) symmetric models combine the best of both worlds.
They introduce right-handed electroweak isospin
in addition to the left-handed isospin of the SM and contain 
a left-triplet $\Delta_L$ with $\vev{\Delta^0_L}=0$ (so that $\rho=1$
is natural) and  a right-triplet $\Delta_R$ with $\vev{\Delta^0_R}\neq 0$
so as to generate a Majorana neutrino mass. L-R symmetry requires
that if the Majorana $\call_M$ is present for $\Delta_R$, 
then it must also be present for $\Delta_L$.
In what follows, we discuss the phenomenology of the $\Delta_L$;
that for the Higgs in $\Delta_R$ is very different \cite{leftright}. We will
drop the $L$ subscript in the following.
Limits on the $h_{ij}$ by virtue of the $\dmm\to
\ell^-\ell^-$ couplings include: Bhabha scattering, $(g-2)_\mu$, 
muonium-antimuonium conversion, and $\mu^-\to e^- e^- e^+$.  
Writing
\begin{equation}
|\hdmm_{\ell\ell}|^2\equiv c_{\ell\ell} \mdmm^2(\gev)\,,
\label{hlimitform}
\end{equation}
$c_{ee}< 10^{-5}$ (Bhabha) and $\sqrt{c_{ee}c_{\mu\mu}}<10^{-7}$
(muonium-antimuonium) are 
the strongest of the limits. There are no limits on $c_{\tau\tau}$.

If $\vevd$ is small or 0, the $\dmm$ width would be quite small,
which can lead to very large $s$-channel production rates
for $\dmm$ in $\emem$ and $\mu^-\mu^-$ collisions \cite{Gunion:1996mq}.
The strategy for $\dmm$ detection is the following.
For small or zero $\vevd$, we would discover the $\dmm$
in $p\anti p,pp \to \dmm\dpp$ with $\dmm \to
\ell^-\ell^-,\dpp\to\ell^+\ell^+$ ($\ell=e,\mu,\tau$) at the Tevatron or LHC
for $\mdmm\lsim 1\tev$ \cite{loomispitts}.
This is precisely the mass range for which $s$-channel
production of the $\dmm$ would be possible at a $\rts\leq 1\tev$
$\emem$ LC or $\mu^-\mu^-$ collider.
Event rates can be enormous for $c_{\ell\ell}$
near the current upper limits \cite{Gunion:1996mq}. 
Equivalently, $\emem$ and $\mummum$
collisions probe very small $c_{\ell\ell}$.
For small beam energy spread ($\delta E/E\equiv R$), 
the number of $\dmm$ produced in $\ell^-\ell^-$ collisions is
\begin{equation}
N(\dmm)_{L=50\fbi}\sim 3\times 10^{10}\left({c_{\ell\ell}\over 10^{-5}}\right)
\left({0.2\%\over R}\right)\,,
\label{eventratenarrow}
\end{equation}
where $R\sim 0.2\%$ is reasonable in $\emem$ collisions
and $R\lsim 0.01\%$ is possible in $\mummum$ collisions.
If 100 events  (of like sign
dilepton pairs of definite known invariant mass) constitute a viable signal, 
Eq.~(\ref{eventratenarrow}) implies we can probe
\begin{equation}
\left.c_{\ell\ell}\right|_{\rm 100~events}\sim 3.3\times 10^{-14}
 \left({R\over0.2\%}\right)
\left({50\fbi\over L}\right)\,,
\end{equation} 
independent of $\mdmm$. This is dramatic sensitivity --- at least a
factor of $10^8-10^9$ improvement over current limits
at an $\emem$ collider. If $\dmm\to \mu^-\mu^-$ primarily, 
then 10 events might constitute a viable signal 
and sensitivity would be further improved.

As a final remark, we note that if triplets are present in a SUSY model,
the triplet Higgs field(s) will destroy coupling constant 
unification if intermediate scale matter is not included;  but, 
this is not a severe problem since such matter is natural in L-R
symmetric supersymmetric models.

\section{Extra Dimensions and Higgs Physics}

This is a very large are of recent research and I will say only
a few words about a variety of interesting possibilities.

The first important point is that large extra dimensions
are associated with much lower Planck scales,
possibly as low as $M_S\sim 1\tev$ \cite{add}.  This 
reduces and can even eliminate the naturalness and hierarchy problems.
In particular, the quadratic divergence in the Higgs mass
loop calculation would be cutoff at $M_S$.
As a result, this particular motivation for low-energy supersymmetry
is greatly reduced.
(Of course, in most such models one must view the MSSM
 unification of gauge couplings at the GUT scale in the usual
four-dimensional theories as being totally accidental.)
Other useful possibilities with large extra dimensions include
various explanations of the small size of most Yukawa couplings.
In one approach \cite{Mirabelli:2000ks}, 
the brane on which the SM particles live
has significant width, and the Higgs is centered at one location
while the weakly coupled fermions are located with significant
separations from the Higgs centrum.
 
Extra dimensions can also provide new contributions to the precision
electroweak observables \cite{edprecision}.  These can shift expectations for
the mass of the SM-like Higgs, in particular allowing it to be
much heavier than the light $\mhsm\sim 100\gev$ values required
in the pure SM context.  Just as in the general 2HDM case, 
the extra dimension theory only needs to give a small $\Delta S$
contribution and a large $\Delta T>0$ contribution.

Extra  dimensions can also be the source of electroweak symmetry breaking.
In one approach \cite{ggewsb}, the Kaluza Klein (KK) modes mix with Higgs
in such a way that the full effective potential takes the form
$\overline V_{\rm tot}=V(\phi)-\overline DV^2(\phi)$, with
$\overline D<0$ from the KK summation. If $\overline D<0$,
as for instance if the number of extra dimension is $\delta=1$,
then the minimum of this potential is at $V(\phi)={1\over 2 \overline D}$,
independent of the form of $V(\phi)$.  In fact, even if $V(\phi)$
has no quartic term, the $-\overline DV^2(\phi)$ term generates the quartic
interactions and EWSB takes place.  The physical Higgs boson is
a complicated mixture of the usual Higgs field and a sum of KK modes.
The main phenomenological implication is that such a Higgs might
not have significant Yukawa couplings and invisible decays into
KK modes could be dominant.

It is also the case that the Lagrangian could contain a term of form
$-{\zeta\over 2}R(g)\phi^\dagger\phi$, where $R(g)$ is the usual
Ricci scalar.  This term introduces mixing between the Higgs
bosons and the KK excitations associated with the extra dimensions.
The result is a large invisible decay width of the Higgs boson
\cite{Giudice:2001av}.

In the Randall-Sundrum model, there is only one graviscalar.
It mixes with the Higgs boson, yielding two mixed physical states
with properties that are intermediate between
those of the radion and of the Higgs boson \cite{Giudice:2001av}.

\section{Detecting an Invisible Higgs Boson}

Aside from invisible KK mode decays, there are also the possibilities
of Higgs decays to Majorans, to $\cnone\cnone$ (in supersymmetric models),
and to 4th generation neutrinos (with $m_{\nu_4}>\mz/2$ to avoid
$Z$ invisible width limits).  If the Higgs decays are dominated
by the invisible channel(s), alternative Higgs detection strategies are
necessary.  At a LC, there is no difficulty in seeing 
an invisibly decay Higgs in Higgsstrahlung production, 
$\epem\to \zstar\to Z\h$,
by looking for a peak in the recoil mass $\mx$ in the $ZX$ final state,
with $Z\to\epem$ and $Z\to \mupmum$.  

At hadron colliders,
detection will be more difficult.   The key is to be
able to tag the Higgs event using particle(s) produced in association
with the Higgs boson.  The modes  
$t\anti t\h$ production \cite{Gunion:1994jf} 
and $W\h,Z\h$ production \cite{Frederiksen:1994me,Choudhury:1994hv} 
were identified
early on as being very promising, but detailed experimental
evaluation/simulation has only recently been begun.
The latter modes might even be useful at the Tevatron \cite{Martin:1999qf}.
More recently, $WW\to\h$ fusion using double
tagging of highly energetic forward jets at the LHC has been proposed 
 \cite{Eboli:2000ze}.
It should be noted that the $W\h$, $Z\h$ and $WW$ fusion 
modes all rely on the $VV$ coupling 
of the Higgs boson, whereas the $t\anti t\h$
mode relies on the fermionic couplings and would be relevant even for
the Higgs bosons of an extended Higgs sector that have small or zero
tree-level $VV$ couplings.  

For a SM-like Higgs, it was estimated that
the $W\h+Z\h$ and $t\anti t\h$ modes have discovery reach
at the LHC up to about $200\gev$ and $250\gev$, respectively, with $L=100\fbi$
of accumulated luminosity. At the Tevatron, the $W\h+Z\h$ modes
will only exceed the limits for an
invisibly decaying SM-like Higgs boson already established
at LEP2 ($\mh>100\gev$) when $L>5\fbi$.
These discovery reaches are substantially less than
those for the $\hsm$ with normal decays.
A roughly equal mixture of invisible and normal decays
would reduce the reach of both the invisible decay and normal decay detection
techniques and  possibly make Higgs
detection all but impossible at the hadron colliders.
A careful study is needed.

\section{Supersymmetric Model Higgs Bosons}

A good summary of the MSSM Higgs sector is found in \cite{hhg}.
At least two doublets are required in supersymmetry
in order to give mass to both up quarks and down quarks and leptons.
An even number of doublets, plus
their higgsino partners, are also required for cancellation of anomalies.
The MSSM contains exactly two doublets ($Y=+1$ and $Y=-1$)
with type-II Yukawa couplings.
TeV scale supersymmetry as embodied in the MSSM
is the most popular cure for the 
naturalness and hierarchy problems for good reason.
First, for two (and only two) doublets one finds
perfect coupling constant unification at the GUT scale if
the SUSY scale is $\msusy\sim 1\tev$ (actually, a significant
number of sparticles with masses nearer $10\tev$ gives better $\alpha_s$
unification with $\alpha_2$ and $\alpha_1$).
If there are more doublets, triplets, etc. then coupling
unification generally requires intermediate
scale matter between the $\tev$ and $\mgut$ scales.
If there are extra dimensions, unification would not necessarily
be relevant (although it can be maintained by putting the SM
particles in the bulk \cite{Dienes:1999vg}). In short, the MSSM without extra
large dimensions has very strong motivation.

The only extension to the MSSM Higgs sector that does not
destroy gauge unification is to add one or more singlet Higgs fields.
The model in which one singlet Higgs field is included is called
the NMSSM (next-to-minimal supersymmetric model)
\cite{Ellis:1989er} (see \cite{hhg} for a review).  This is an
extremely attractive model in that it provides the most
natural explanation for having a $\mu$ parameter that has
TeV scale magnitude. The parameter $\mu$ is that appearing in
the MSSM superpotential $\mu \what \phi_u \what \phi_d$,
where the $\what \phi_{u,d}$ are the superfields containing
the scalar $\phi_{u,d}$ Higgs fields of the type-II Higgs sector.
In the NMSSM, this interaction is replaced by the superpotential
form $\lam_S \what S \what\phi_u\what \phi_d$, which generates
an electroweak scale effective $\mu=\lam_S s$ 
when the scalar component of $\what S$
acquires an electroweak scale vacuum expectation value, 
$s\equiv \vev{S^0}$, as is
easily and naturally arranged. In addition, the NMSSM
can contain a superpotential piece of the form ${1\over 3}\kappa \what S^3$.

As is well known, there is a strong bound on the mass of the lightest
Higgs boson $\hl$ of supersymmetric models. In the MSSM,
if $\mstop\leq 1\tev$ then 
$\mhl\lsim 130-135\gev$ after including stop loop corrections
with $A_t\neq 0$. ($A_t$ is the magnitude of the trilinear
soft supersymmetry breaking term.) This bound is so strong
because {\it at tree-level} one finds $\mhl\leq\mz$
due to the fact that all Higgs self couplings are given in terms
of gauge couplings, $g$ and $g'$.
 However, the choice above of $\mstop\leq 1\tev$
is a bit arbitrary. As noted earlier, having some SUSY matter
nearer $10\tev$ actually improves coupling constant unification.
For stop masses in this latter range, the upper bound on $\mhl$ 
would be larger.
Also, increasing the top mass within the current experimental
error increases the upper limit on $\mhl$.
In the NMSSM, the upper bound is less constrained because
of the new $\lam_S$ parameter introduced. One finds
$\mhl\leq 150\gev$ {\it assuming perturbativity
for $\lam_S$ up to $\mgut$.} If one adds more doublet Higgs
superfields, this
actually lowers the mass bound. Adding triplet Higgs superfields
increases the mass bound (assuming perturbativity up to $\mgut$ again) to 
$\mhl\leq 200\gev$ \cite{Espinosa:1993hp}.
This is the maximal value employed earlier in the sum rule of 
Eq.~(\ref{msqlim}).

\subsection{Experimental limits from LEP2 on MSSM Higgs bosons}

Limits from LEP2 in the MSSM context are quite significant.
Roughly, $\mhl,\mha\lsim 91\gev$ are excluded \cite{LEPHIGGSgroup} for 
maximal mixing in the stop squark sector
(a certain choice of $X_t\equiv A_t-\mu\cot\beta$)
and $\msusy=1\tev$. Using the
theoretical upper bound on $\mhl$ as a function of $\tanb$,
this translates to exclusion of the region
$0.5<\tanb<2.4$ at 95\%~CL.  
Higher $\msusy$ means that
Higgs masses at a given $\tanb$ increase with the result
that less of the $[\mha,\tanb]$ parameter space
is excluded. 

The above limits on $\mhl,\mha$ assume absence of CP violation
in the Higgs sector and invisible decays of the $\hl,\ha$
are not allowed for.
CP violation arises in the MSSM through phases of the $\mu$
and $A_t$ parameters.
This CP violation leads to CP violation in the MSSM two-doublet Higgs
sector brought in via the one-loop corrections sensitive to these phases.
The two new parameters are: $\phi_\mu+\phi_A$ and
$\theta$, the latter being the phase of one of the Higgs doublet fields
relative to the other.
Studies \cite{Kane:2000aq,Carena:2000ks} suggest that MSSM Higgs mass
limits will be weakened significantly, implying that the disallowed
$\tanb$ region is still allowed when CP violation is present.

Allowing for $\hl$ and $\ha$ to have some, 
perhaps substantial, invisible decays would 
considerably weaken the constraints on
the $\hl\ha$ cross section. As a result,
the $ZX$ recoil mass analysis would have to be relied upon more heavily.
I would guess that the limits on $\mhl$ and $\mha$ deteriorate substantially.
This deserves study by the LEP experimental groups.

\subsection{Discovery prospects for MSSM Bosons at the Tevatron} 

\begin{figure}[h!]
\centering
\centerline{\psfig{file=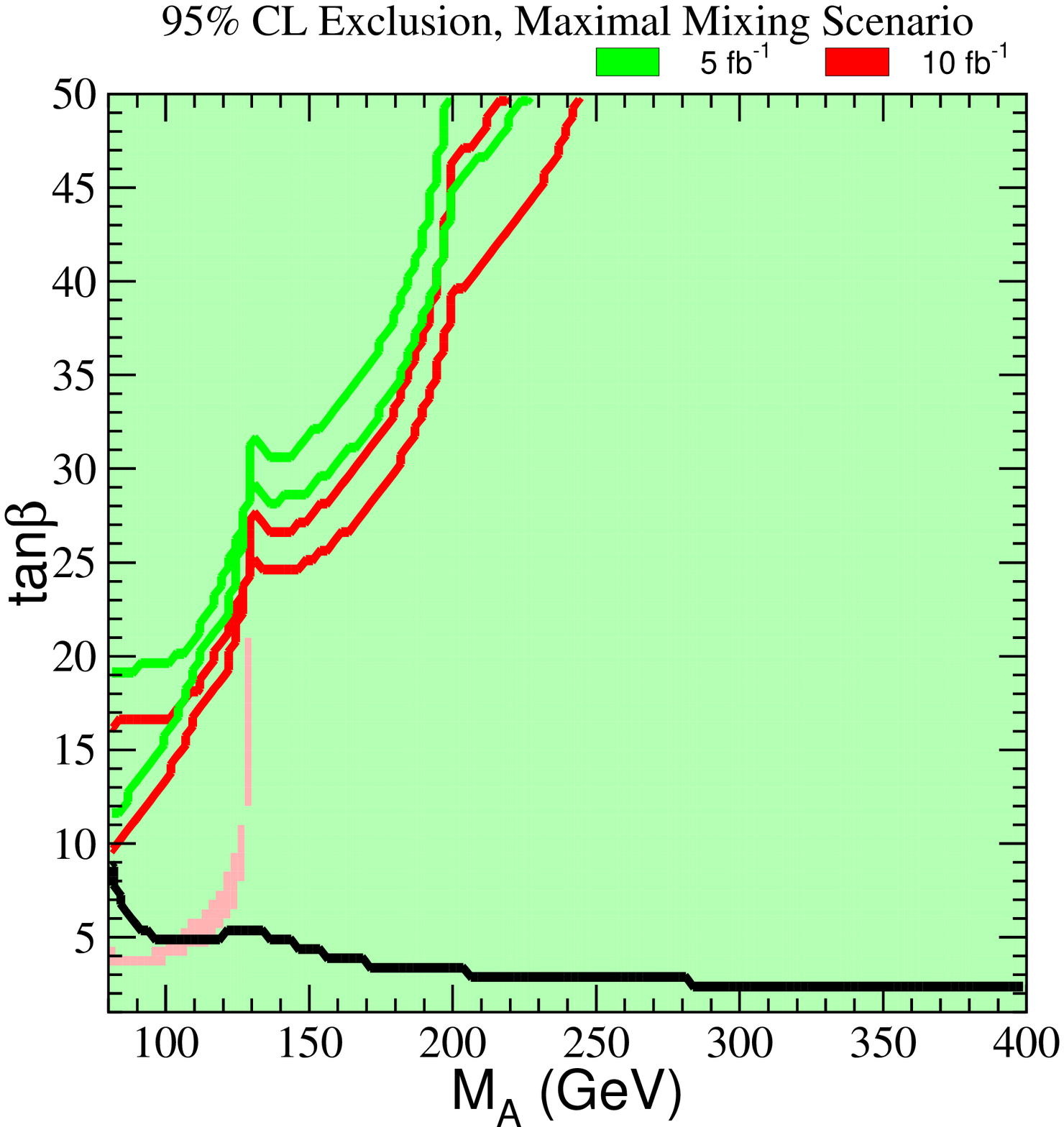,width=8cm}
\psfig{file=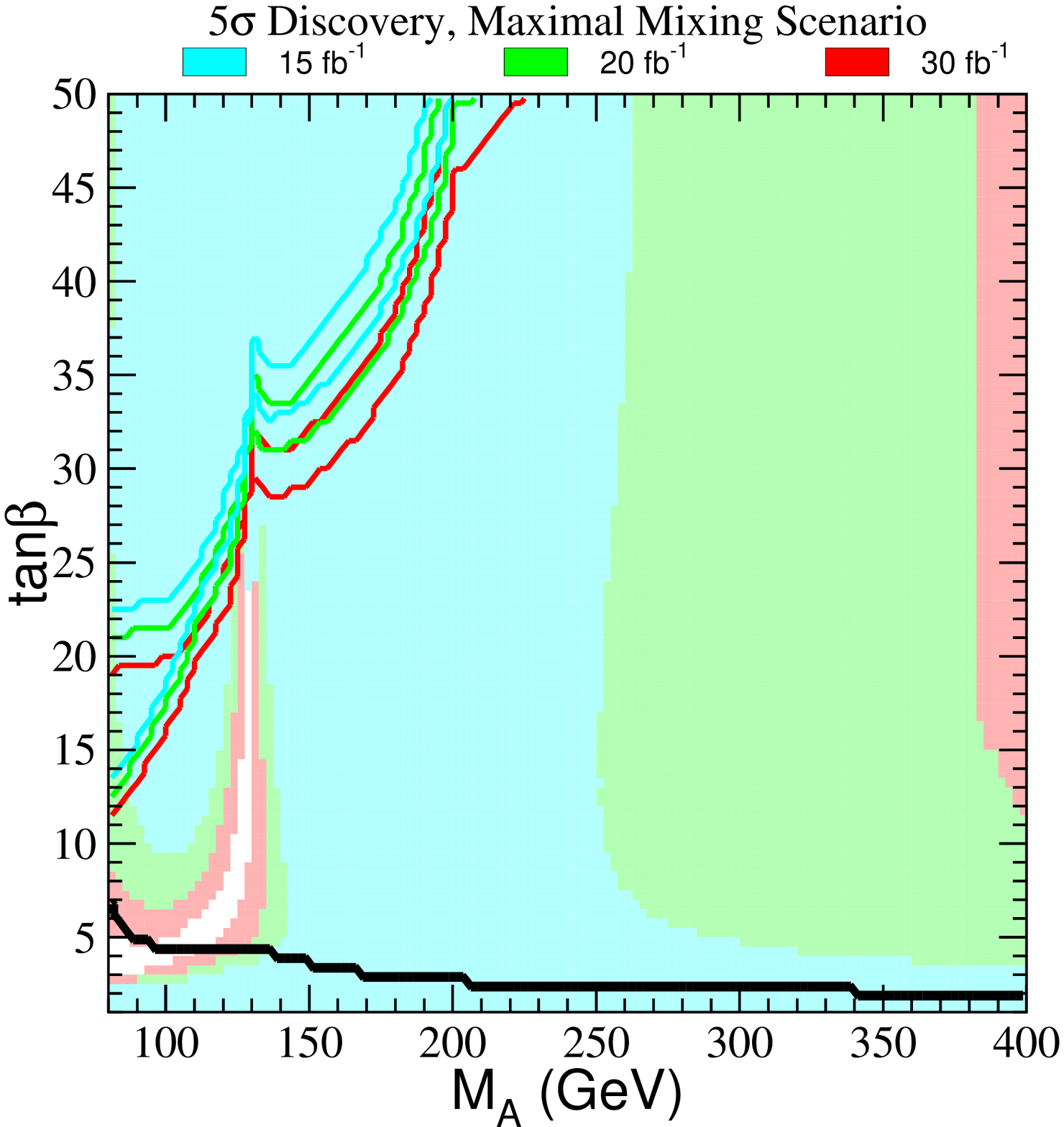,width=8cm}}
\caption[0]{\label{fullmhmax95} (a) 95$\%$ CL exclusion regions
and (b) $5\sigma$ discovery regions in the $[\mha,\tan \beta]$
plane, for the maximal mixing scenario (using $\msusy=1\tev$)
and two different search channels:
$q\bar q\to V\phi$ [$\phi=\hl$, $\hh$], $\phi\to b\bar b$
(shaded regions) and 
$gg$, $q\bar q\to b\bar b\phi$ [$\phi=\hl$, $\hh$, $\ha$],
$\phi\to b\bar b$ (region in the upper left-hand corner bounded by the
solid lines).  
The region below the solid black line is excluded by no $e^+e^-\to Z\phi$
events at LEP2.
} \label{tevsum}
\end{figure}

We recall that the $\hh$ ($\hl$) has most of the $WW,ZZ$ coupling 
when $\mha\lsim \mz$ ($\mha\gsim 150\gev$). For, $\mz\lsim\mha\lsim 150\gev$,
the $\hl$ and $\hh$ will share the $WW,ZZ$ coupling strength to a greater
or lesser extent depending upon other details of the input parameters.
The useful production processes are
$q\bar q \to V \hl,V\hh$  with $\hl,\hh\to b\anti b$ being dominant
for $\mha$ values such that $\hl,\hh$ has substantial $VV$
coupling, respectively.  The decoupled Higgs boson, $\hl$ at low $\mha$ or
$\hh$ at high $\mha$, will have $b\anti b$ coupling that is
enhanced by a factor of $\tanb$
(relative to the SM-like value) and
the processes $gg,q\bar q\to b\bar b\hl$ or $b\bar b\hh$, respectively,
will be enhanced; $gg,q\bar q\to b\anti b\ha$ is always enhanced at high
$\tanb$. Obviously, a Higgs which decouples from $VV$ and has enhanced
$b\anti b$ coupling will decay primarily to $b\anti b$. A careful
study, including a parameterized simulation of detector effects, was
performed to study prospects at the Tevatron using
these channels \cite{tevreport}.  Except for some very special parameter
configurations, the results are summarized by Fig.~\ref{tevsum}.
One sees that $L>15\fbi$ is needed to guarantee
$5\sigma$ discovery at lower $\mha$. For larger $\mha$,
as typically needed for successful generation of EWSB
via the RGE running, much higher $L$ will be needed.
Except in the upper left corner of low $\mha$ and high $\tanb$,
only the $\hl$ is observed. The small white region is where
the $\hl$ and $\hh$ are sharing the $VV$ coupling and neither
is produced strongly enough for detection. If the root-mean squark
mass, $\msusy$, is increased above $1\tev$, the $\hl$ mass
increases and discovery prospects deteriorate.

\subsection{Discovery Prospects for MSSM Higgs Bosons at the LHC}

\begin{figure}[t!]
\vspace{-.5in}
\centering
\psfig{file=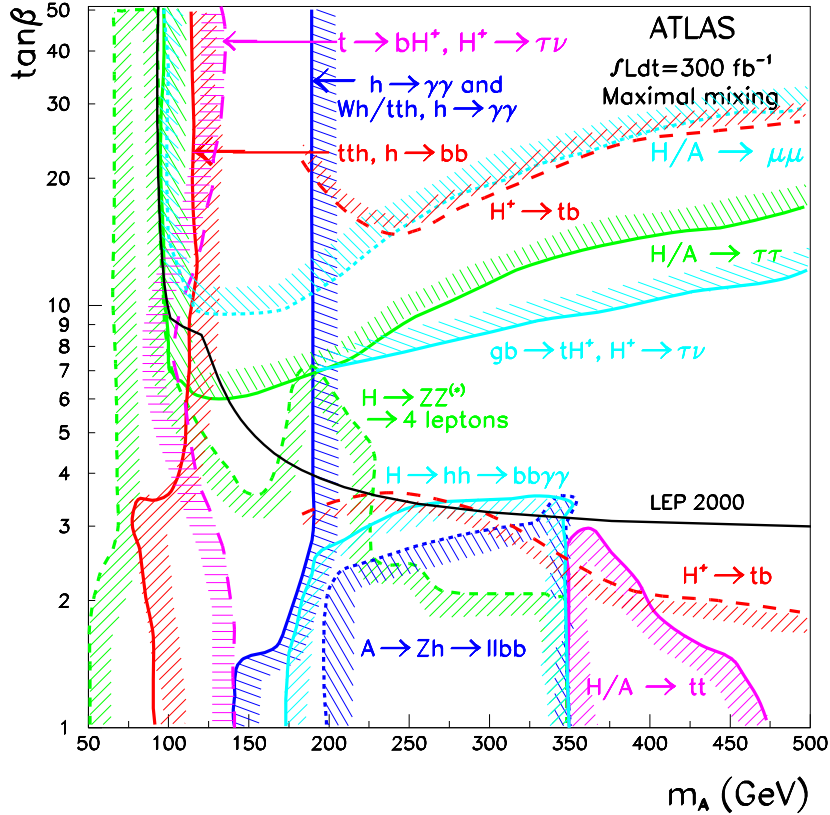,width=12cm}
\caption[0]{$5\sigma$ discovery contours for MSSM Higgs boson detection
in various channels are shown in the $[\mha,\tanb]$ parameter plane, 
assuming maximal mixing, $\msusy=1\tev$
 and an integrated luminosity of $L=300\fbi$
for the ATLAS detector. This figure is preliminary \cite{atlasmaxmix}.}
  \label{atlasmssm}
\end{figure}

Focusing on large $\mha$,  discovery of the SM-like $\hl$  
will typically be possible using 
the same production/decay modes as for a light $\hsm$ at the LHC.
At high $\tanb$ and large $\mha$, the decoupled $\hh$ and $\ha$
can be found using 
$gg,q\bar q\to b\bar b\hh,b\bar b \ha$, with $\hh,\ha\to
\tauptaum$ or $\mupmum$ and $gb\to \hpm t$ with 
$\hpm\to \tau^\pm \nu$. These are the main modes of importance
since LEP2 limits pretty much exclude $\tanb<2.5$, for which other modes 
could be dominant. The contours for $5\sigma$ discovery are shown in
Fig.~\ref{atlasmssm}. Discovery of at least one of the MSSM Higgs bosons
is guaranteed for $L=300\fbi$.  If $\mha\gsim 200\gev$
and $\tanb$ is not large enough,
it could happen that only the SM-like $\hl$ will be observable. 

\subsection{Discovery Prospects for MSSM Higgs Bosons at the LC}

Recent reviews of this topic include \cite{teslareport} and
\cite{orangebook}.
Any Higgs boson with even very modest $VV$ coupling can be
detected using the Higgsstrahlung $\epem\to \zstar\to Z\h$
process. For $\mha\gsim 150\gev$ (as probable for
RGE driven EWSB), decoupling has set in and
it is the $\hl$ that will be detected in this way.
In particular, the upper limit of $\mhl\lsim 135\gev$ guarantees
that a $\rts=350\gev$ LC would suffice.
For the $\hh,\ha,\hpm$, the 
production mechanisms $\epem\to \hh+\ha$ and $\epem\to\hp+\hm$ would
be full strength. However, at large $\mha$, one finds
$\mhh\sim\mha\sim\mhpm$ so that pair production requires
$\mha<\rts/2$. If $\mha$ exceeds $\rts/2$, then one must turn to 
$\epem\to b\anti b \ha,b\anti b \hh,
bt\hpm$. As we have already discussed, the event rates for these
processes are not large enough for observation unless $\tanb$
is quite large.  In the problematical moderate $\tanb$ wedge,
where the LHC will also not find the $\hh,\ha,\hpm$, observation might
be possible using $\gam\gam\to \hh,\ha$. In particular, this will be
possible if the value of $\mha$ is constrained to within $\pm 50\gev$,
since then the expected yearly luminosity would be such that an
appropriately designed scan,
using a peaked $\gam\gam$ luminosity spectrum, would reveal a $\hh,\ha$
signal when $E_{\gam\gam}^{\rm peak}\sim \mha$  \cite{twogamhiggs}.

A model-dependent constraint on $\mha$ of this type might be possible.
If one assumes a non-conspiratorial MSSM parameter scenario, 
$\hl$ vs. $\hsm$ branching ratio differences reflect the value of $\mha$
rather accurately. Further, expected LC precisions for the
branching ratios are such that these differences
could be measured with sufficient accuracy to determine 
$\mha$ within $\pm 50\gev$ if $\mha\lsim 400\gev$
\cite{Gunion:1996cn,deviations}. This is precisely the range
of relevance for a $\gam\gam$ collisions at a $\rts=500\gev$ LC.
Alternatively, if the properties of the observed light Higgs
are found not to deviate from those predicted for the $\hsm$,
then the most natural conclusion would be that $\mha$ is substantially
heavier than $500\gev$. In this case, a $\gam\gam$ scan 
over the $E_{\gam\gam}\lsim 400\gev$ range would not be useful.
However, if one does not accept this model-dependent
indirect determination of the magnitude of $\mha$, a full scan,
say from $\mha\sim 250\gev$ up to $\mha\sim 400\gev$ would
be called for. However, luminosity expectations for the NLC design might
not suffice \cite{twogamhiggs} to find the $\hh,\ha$ if
one has to scan such a large range of mass. 
Much higher $L_{\gam\gam}$ luminosity is claimed by
TESLA. This might be a rather
crucial difference \cite{twogamhiggs,Muhlleitner:2001kw}. 
Once the mass of any of the $\hl,\hh,\ha$ is known, we 
can run with $E_{\gam\gam}^{\rm peak}$ equal to the Higgs mass and determine
the CP nature of the Higgs boson by
adjusting the linear polarization orientations of the
initial laser beams \cite{Grzadkowski:1992sa,Gunion:1994wy,Kramer:1994jn}.  
In particular, we can separate
$\ha$ from $\hh$ when these are closely degenerate (as typical for
$\tanb\gsim 4$ and $\mha>2\mz$).

\subsection{Special Cases in the MSSM}

As already noted, the above summaries assume relatively canonical
MSSM parameter choices and absence of CP violation in the Higgs
sector. These expectations need not apply.
If there are substantial 
$\hl\to\cnone\cnone$ decays, as still possible even given LEP2 
lower bounds on $\mcnone$, observation of the $\hl$
at hadron colliders (but not the LC) would be more difficult.
For low stop masses, corrections to the one-loop induced $gg\hl$
and $\gam\gam\hl$ couplings would be substantial.
The stop and top loops negatively interfere leading to
reduction of $gg$ fusion production and some
increase in $\br(\hl\to\gam\gam)$ \cite{hhg,Djouadi:1998az}.

There can be substantial radiative corrections to the tree-level couplings.
This would be especially important for $b\anti b$ decays of 
the $\hl$ when the $\hl$ is SM-like. In particular, after including
radiative corrections, for the $b\anti b$ Yukawa Lagrangian one obtains 
$
{\cal L} \simeq \lam_b \phi_d^0 b \anti{b} + \Delta \lam_b \phi_u^0 b \anti{b}.
$
The coupling $\Delta \lam_b$ is one-loop and arises from
$\sbot-\gl$ and $\stop - \wtil \phi_{u,d}$ loops.
 Typically, ${\Delta\lam_b\over\lam_b}\sim 0.01$ (either sign).
Further,
${\Delta\lam_b\over\lam_b}$ does not vanish in the limit of large SUSY masses 
(there is no decoupling).
 The result for the full $\hl\to b\anti b$ 
coupling takes the form:
\begin{equation}
\lam_b^{\hl}  \simeq -\frac{m_b \sin\alpha}{v \cos\beta}\frac{1}{1+{\Delta\lam_b\over\lam_b}\tanb}
\left[ 1 - \frac{{\Delta\lam_b\over\lam_b}}
{\tan\alpha} \right],
\label{barhb}
\end{equation}
implying that if $\tan\alpha \simeq \frac{\Delta\lam_b}{\lam_b}$
then $\lam_b^{\hl}\simeq 0\,.$ In particular, this can
happen when $\mha\to \infty$ if $\Delta\lam_b/\lam_b<0$,
since, at large $\mha$, $\alpha\to \pi/2-\beta$ and 
$\tan\alpha\to -1/\tanb$ is small. Conversely, for $\Delta\lam_b/\lam_b>0$,
substantial enhancement of $\lam_b^{\hl}$ is possible.

If parameters are such that 
the $\hl$ decouples from $b$'s (i.e. $\hl\simeq \Re\phi_u^0$),
discovery strategies could not rely on the $b\anti b$ decay mode.
However, since $\Gamma(\hl\to\gam\gam)$ is dominated by $W$ and $t$ loops,
small or vanishing $\lam_b^{\hl}$ will affect the $\hl\to\gam\gam$ partial width
very little. There is also little impact on the $gg$ partial width.
Thus, suppressed 
$\Gamma(\hl\to b\anti b)$ implies enhanced $\br(\hl\to \gam\gam),
\br(\hl\to WW^*)$. In fact, 
the $\gam\gam$ mode can be viable for some range of
$\mhl$ at the Tevatron if $\hl\sim \phi_u$ \cite{Mrenna:2001qh}.
More generally, allowing for either suppressed or enhanced $\lam_b^{\hl}$,
LHC $gg\to \hl\to\gam\gam$ and Tevatron $W\hl[\to WW^*]$ 
modes improve when LHC, Tevatron $W,Z\hl[\to b\anti b]$ modes deteriorate.
One also finds that the Tevatron and the LHC are complementary 
as $\lam_b^{\hl}$ and $\mhl$ vary in 
that $\hl$ discovery will occur at one or the other machine,
even if not both \cite{Carena:2000bh}.

Turning next to the  
$\hh,\ha,\hpm$, discovery will typically become more difficult
if these Higgs bosons have substantial branching ratios
for decay to pairs of neutralinos, or charginos
or sleptons, $\ldots$.
Such decays will, however, only be significant if $\tanb$ is in
the low to moderate range, a significant part of which has already
been excluded by LEP2 data. For larger $\tanb$, the $\hh,\ha\to b\anti b$
and $\hp\to t\anti b$ 
decay modes and their $\tau$ analogues
are sufficiently enhanced that sparticle pair channels will
have small branching ratios.

\subsection{Discovery of NMSSM Higgs Bosons}

The addition of the singlet superfield
results in a third CP-even Higgs boson and a second CP-odd Higgs
boson. The CP-even bosons mix, as do the CP-odd bosons.
There is still a strong constraint of $m_{h_1^0}\leq 150\gev$ on the mass
of the lightest CP-even physical state. If it does not have substantial
coupling to $VV$, then it can be shown that one of the other two
states ($h_2^0$ or $h_3^0$) 
must have at least moderate $VV$ coupling {\it and}
must be relatively light.  
As a result, discovery of one (or more) of the CP-even
Higgs bosons of the NMSSM is guaranteed at a LC with 
$\rts>350\gev$ \cite{Ellwanger:1999ji}.
An important question is whether
the sharing of the $VV$ coupling that is possible
in the NMSSM means that discovery of one of the NMSSM Higgs bosons
at the LHC cannot be guaranteed. A study for Snowmass 96 \cite{Gunion:1996fb}
showed that parameters could be chosen so that no Higgs
boson would be observed  employing the experimentally
verified modes available at that time, even for $L=600\fbi$.
For example, for $\mhi=105\gev$ and $\tanb=5$, event rates in
the $h_{1,2,3}\to \gam\gam$, $Z\zstar$, $W\wstar$, \etc\ final states
could all be suppressed by virtue of a shared $VV$ coupling
configuration, while $\tanb$ was too small to sufficiently enhance
the $b\anti b h_{1,2,3}$ and $b\anti b a_{1,2}$ 
(with $h_{1,2,3},a_{1,2}\to \tauptaum$) modes
to the $5\sigma$ level.  What was missing in 1996 was a discovery
mode based on the $b\anti b$ decays, especially those of the lightest 
Higgs bosons, $h_{1,2}$.
Recently, the $t\anti t \h\to t\anti t b\anti b$ mode 
(originally discussed in \cite{Dai:1993gm})
has been shown to be viable for a SM-like Higgs
by the ATLAS and CMS groups \cite{lhctdr}. Rescaling these results
to the NMSSM, a preliminary study \cite{gunell} finds that all points 
for which discovery was found to be impossible without this mode
would allow $>5\sigma$ discovery of $\hi$ or $\hii$ in the
$t\anti t b\anti b$ final state. Further study is needed,
but it now appears that there is a `no-lose' theorem
for NMSSM Higgs discovery at the LHC once both ATLAS and CMS
have each accumulated $L\geq300\fbi$ of luminosity.

\section{Conclusions}

This brief overview of discovery prospects for Higgs bosons necessarily
omitted many interesting topics, and almost completely ignored the
very interesting programs for precision measurements of the 
properties of the Higgs bosons at the LHC and LC and how such
measurements impact our ability to determine, for instance,
the MSSM or NMSSM soft-supersymmetry-breaking parameters.
It is these latter which are needed to connect
TeV scale physics to the GUT scale physics that we ultimately hope to probe.

\section*{Acknowledgments}
This work was supported in part by the U.S. Department of Energy
and by the Davis Institute for High Energy Physics.

\end{document}

%% file: econfmacros2.tex
\def\beq{\begin{equation}}
\def\eeq#1{\label{#1}\end{equation}}
\def\eeqn{\end{equation}}

\newenvironment{Eqnarray}%
   {\arraycolsep 0.14em\begin{eqnarray}}{\end{eqnarray}}
\def\beqa{\begin{Eqnarray}}
\def\eeqa#1{\label{#1}\end{Eqnarray}}
\def\eeqan{\end{Eqnarray}}

\let\bar=\overbar

\def\ie{{\it i.e.}}

\def\vev#1{\langle #1 \rangle}

\def\Dslash{\not{\hbox{\kern-4pt $D$}}}
\def\dslash{\not{\hbox{\kern-2pt $\del$}}}

\def\mz{m_Z}

\def\mw{m_W}

\def\mh{m_h}

\def\msb{{\bar{\ssstyle M \kern -1pt S}}}

\def\eps{\epsilon}

\def\lsim{\mathrel{\raise.3ex\hbox{$<$\kern-.75em\lower1ex\hbox{$\sim$}}}}
\def\gsim{\mathrel{\raise.3ex\hbox{$>$\kern-.75em\lower1ex\hbox{$\sim$}}}}